\documentclass{emulateapj}
\usepackage{graphicx}
\usepackage{txfonts}
\usepackage{natbib}
\usepackage[scaled]{helvet}
\usepackage{epsfig}
\usepackage{url}
\bibpunct{(}{)}{;}{a}{}{,}
\newcommand{\kms}{\,km\,s$^{-1}$}
\usepackage{textcomp}
\usepackage{mathpazo}

\usepackage{color}
\usepackage[normalem]{ulem}

\newcommand{\teff}{\ensuremath{T_{\rm eff}}}

\newcommand{\msun}{\ensuremath{\,M_\Sun}}
\newcommand{\rsun}{\ensuremath{\,R_\Sun}}

\newcommand{\ms}{\,m\,s$^{-1}$}
\newcommand{\minus}{\scalebox{0.75}[1.0]{$-$}}

\begin{document}

\title{DM Ori: A Young Star Occulted by a Disturbance in its Protoplanetary Disk}
%\kgsdel{Another Young Star with a Disk Goes ``Bump'' in the Night}
\author{Joseph E. Rodriguez$^1$, Keivan G. Stassun$^{1,2}$, Phillip Cargile$^{3}$, Benjamin J. Shappee$^{4,5}$, Robert J. Siverd$^{6}$, Joshua Pepper$^{7}$, Michael B. Lund$^1$, Christopher S. Kochanek$^{8,9}$, David James$^{10}$, Rudolf B. Kuhn$^{11}$, Thomas G. Beatty$^{12,13}$, B. Scott Gaudi$^{8}$, David A. Weintraub$^{1}$, Krzysztof Z. Stanek$^{8,9}$, Thomas W.-S. Holoien$^{8,9,14}$, Jose L. Prieto$^{15,16}$, Daniel M. Feldman$^{17}$, Catherine C. Espaillat$^{17}$}

\affil{$^1$Department of Physics and Astronomy, Vanderbilt University, 6301 Stevenson Center, Nashville, TN 37235, USA}
\affil{$^2$Department of Physics, Fisk University, 1000 17th Avenue North, Nashville, TN 37208, USA}
\affil{$^3$Harvard-Smithsonian Center for Astrophysics, 60 Garden St, Cambridge, MA 02138, USA}
\affil{$^{4}$Carnegie Observatories, 813 Santa Barbara Street, Pasadena, CA 91101, USA}
\affil{$^{5}$Hubble, Carnegie-Princeton Fellow}
\affil{$^6$Las Cumbres Observatory Global Telescope Network, 6740 Cortona Dr., Suite 102, Santa Barbara, CA 93117, USA}
\affil{$^{7}$Department of Physics, Lehigh University, 16 Memorial Drive East, Bethlehem, PA 18015, USA}
\affil{$^{8}$Department of Astronomy, The Ohio State University, Columbus, OH 43210, USA}
\affil{$^{9}$Center for Cosmology and AstroParticle Physics (CCAPP), The Ohio State University, 191 W.\ Woodruff Ave., Columbus, OH 43210, USA}
\affil{$^{10}$Cerro Tololo InterAmerican Observatory, Casilla 603, La Serena, Chile}
\affil{$^{11}$South African Astronomical Observatory, PO Box 9, Observatory 7935, South Africa}
\affil{$^{12}$Department of Astronomy \& Astrophysics, The Pennsylvania State University, 525 Davey Lab, University Park, PA 16802}
\affil{$^{13}$Center for Exoplanets and Habitable Worlds, The Pennsylvania State University, 525 Davey Lab, University Park, PA 16802}
\affil{$^{14}$US Department of Energy Computational Science Graduate Fellow}
\affil{$^{15}$N«ucleo de Astronom«õa de la Facultad de Ingenier«õa, Universidad Diego Portales, Av. Ej«ercito 441, Santiago, Chile}
\affil{$^{16}$Millennium Institute of Astrophysics, Santiago, Chile}
\affil{$^{17}$Department of Astronomy, Boston University, 725 Commonwealth Avenue, Boston, MA 02215, USA}

\shorttitle{DM Ori}

\begin{abstract}
%\kgsdel{One way to better understand the differences in the large variety of planetary systems is to observe the rare occasion when a young star is eclipsed by a component of its circumstellar environment. These ``Disk Eclipsing'' systems allow for a direct measurement of the occulting material, typically from the star's circumstellar disk. By studying a variety of these systems, at different stages in their evolution, we can better understand how these systems eventually evolve into exoplanetary systems.}
In some planet formation theories, protoplanets grow gravitationally within a young star's protoplanetary disk, a signature of which may be a localized disturbance in the disk's radial and/or vertical structure. Using time-series photometric observations by the Kilodegree Extremely Little Telescope South (KELT-South) project and the All-Sky Automated Survey for SuperNovae (ASAS-SN), combined with archival observations, we present the discovery of two extended dimming events of the young star, DM Ori. This young system faded by $\sim$1.5 mag from 2000 March to 2002 August and then again in 2013 January until 2014 September (depth $\sim$1.7 mag). We constrain the duration of the 2000-2002 dimming to be $<$\ 860 days, and the event in 2013-2014 to be $<$\ 585 days, separated by $\sim$12.5 years. A model of the spectral energy distribution (SED) indicates a large infrared excess consistent with an extensive circumstellar disk. Using basic kinematic arguments, we propose that DM Ori is likely being periodically occulted by a feature (possibly a warp or perturbation) in its circumstellar disk. In this scenario, the occulting feature is located $>$6 AU from the host star, moving at $\sim$14.6 \kms, and is $\sim$4.9 AU in width. This localized structure may indicate a disturbance such as may be caused by a protoplanet early in its formation.
\end{abstract}


\section{Introduction}
Although planets form from protoplanetary disks of gas and dust, we do not yet know which aspects of a star's circumstellar environment govern the architectural differences seen in the thousands of planetary systems discovered. Studying a variety of protoplanetary disks at a variety of ages (or planetary formation stages) in great detail may allow us to disentangle the causes for the differences in planetary architecture. One unique way to better understand these differences is to observe the rare occasion when a young star is eclipsed by a component of its circumstellar environment. These dimming events can last many months to years, and can dim the system by $>$ 1 mag \citep{Carroll:1991, Bouvier:2013, Rodriguez:2013}. Indeed, such disk occultations may signal regions of the disk wherein a protoplanet is forming and locally disturbing the disk.

\begin{table*}
\centering
\caption{Stellar Properties of DM Ori obtained from the literature.}
\label{tbl:Host_Lit_Props}
\begin{tabular}{llccc}
   \hline
  \hline
\hline
  Parameter & Description & Value & Source & Reference(s) \\
Names& 					&  DM Ori	& 		&			\\
			& 					& TYC 4775-139-1		& 		&			\\
			&					&2MASS J05433900-0504029&		&			\\
			&					&				&		&			\\
$\alpha_{J2000}$	&Right Ascension (RA)& 05:43:39.0			& Tycho-2	& \citet{Hog:2000}	\\
$\delta_{J2000}$	&Declination (Dec)& $\minus$05:04:02.83			& Tycho-2	& \citet{Hog:2000}	\\
B$_T$			&Tycho B$_T$ magnitude& 12.28 $\pm$ 0.15		& Tycho-2	& \citet{Hog:2000}	\\
V$_T$			&Tycho V$_T$ magnitude& 11.72 $\pm$ 0.12		& Tycho-2	& \citet{Hog:2000}	\\
			&					&				&		&			\\
$V$ & Johnson V&11.545 $\pm$ 0.066& APASS 	& \citet{Henden:2015}	\\
$B$ &Johnson B&12.499  $\pm$  0.026& APASS 	& \citet{Henden:2015}	\\
$g'$&   Sloan g'	&11.972  $\pm$  0.025& APASS 	& \citet{Henden:2015}	\\
$r'$ &  Sloan r'	&11.215  $\pm$  0.043& APASS 	& \citet{Henden:2015}	\\
$i'$  &Sloan i'	&10.924 $\pm$  0.054& APASS 	& \citet{Henden:2015}	\\
			&					&				&		&			\\
J			&2MASS magnitude& 10.211  $\pm$ 0.024		& 2MASS 	& \citet{Cutri:2003}	\\
H			&2MASS magnitude& 9.152 $\pm$ 0.024	& 2MASS 	& \citet{Cutri:2003}	\\
K			&2MASS magnitude& 8.245  $\pm$ 0.023		& 2MASS 	& \citet{Cutri:2003}	\\
			&					&				&           &			\\
\textit{WISE1}		&WISE passband& 6.997  $\pm$ 0.034		& WISE 		&\citet{Cutri:2012}	\\
\textit{WISE2}		&WISE passband&  6.270  $\pm$ 0.021		& WISE 		& \citet{Cutri:2012}\\
\textit{WISE3}		&WISE passband& 3.757 $\pm$ 0.013		& WISE 		& \citet{Cutri:2012}	\\
\textit{WISE4}		&WISE passband& 1.556 $\pm$ 0.018 &WISE 		& \citet{Cutri:2012}	\\
			&					&				&		&			\\
IRAS 12$\mu$m			&IRAS Flux Density (Jy)&1.62 $\pm$ 0.15 &IRAS&		\citet{Helou:1988}	\\
IRAS 25$\mu$m			&IRAS Flux Density (Jy)&2.87 $\pm$ 0.23				&IRAS&\citet{Helou:1988}\\
IRAS 60$\mu$m			&IRAS Flux Density (Jy)&	2.75 $\pm$ 0.44			&IRAS&\citet{Helou:1988}\\
IRAS 100$\mu$m			&IRAS Flux Density (Jy)&	{\bf 2.83}		&IRAS&\citet{Helou:1988}\\
			&					&				&		&			\\
$\mu_{\alpha}$		& Proper Motion in RA (mas yr$^{-1}$)	& 3.7$\pm$ 2.6	& NOMAD		& \citet{Zacharias:2004} \\
$\mu_{\delta}$		& Proper Motion in DEC (mas yr$^{-1}$)	&  $\minus$2.3 $\pm$ 1.9	& NOMAD		& \citet{Zacharias:2004} \\
			&					&				&		&			\\
\hline
\hline
\end{tabular}

 \footnotesize \textbf{\textsc{NOTES}}\\
 \footnotesize Bold values correspond to upper limits (S/N $<$ 2)

%\footnotesize $^{*}$U is positive in the direction of the Galactic Center 
\end{table*}

Using time series photometric observations from the Kilodegree Extremely Little Telescope (KELT) survey, we have been conducting the Disk Eclipse Search with KELT (DESK) survey \citep{Rodriguez:DESK} to look for these rare but valuable events. Through this survey, we have discovered and/or analyzed ``Disk Eclipsing'' events around the young stars RW Aurigae, V409 Tau, and AA Tau \citep{Rodriguez:2013, Rodriguez:2015, Rodriguez:2016A}. In addition, we have found the longest period eclipsing binary ($\sim$69 years), TYC 2505-672-1, an M giant being eclipsed by a hot, optically dim companion surrounded by a disk a few AU in diameter \citep{Rodriguez:2016B}. Studying these large dimming events caused by circumstellar material allows us to study the evolution of a young stellar object's (YSO's) circumstellar environment.

%{\bf [Joey, I'm not understanding the purpose of this entire paragraph. What is the point?]} 
One possible type of disk eclipsing YSOs are UX Orionis stars (UXors), pre-main sequence stars that exhibit large photometric dimming events with durations of months to years and depths $<$3 magnitudes \citep{Grinin:1991}. Their spectral energy distributions (SEDs) imply that they are surrounded by large circumstellar disks. One signature feature of the large dimming events of UXor stars is a reversal of the observed color, from red to blue during the maximum depth of the event. The leading explanations for these dimmings are dust clumps in a nearly edge-on disk occulting the host star \citep{Wenzel:1969, Grinin:1988, Voshchinnikov:1989, Grinin:1998, Grady:2000} and hydrodynamical fluctuations of the inner disk \citep{Dullemond:2003}. Typically, UXors are higher mass Herbig Ae/Be stars, but a few late-type stars (G-type or later) are known to show similar behavior. Two examples of late-type UXor stars are AA Tau (K7, \citealt{Bouvier:1999}) and V409 Tau (M1.5, \citealt{Luhman:2009}). Starting in 2011 and continuing to this day, AA Tau has been fainter than its typical brightness by $\sim$2 mag \citep{Bouvier:2013}. V409 Tau showed two $\sim$600 day long, $\sim$1.4 mag deep dimming events in early 2009 and early 2012 \citep{Rodriguez:2015}. It has been proposed for both systems that a feature, such as a warp or dust clump, in Keplerian orbit is responsible for the large observed dimmings \citep{Bouvier:2013, Rodriguez:2015}. The signature color reversal of UXor systems has been observed for AA Tau but not for V409 Tau. 

\begin{figure*}[!ht]
\centering\epsfig{file=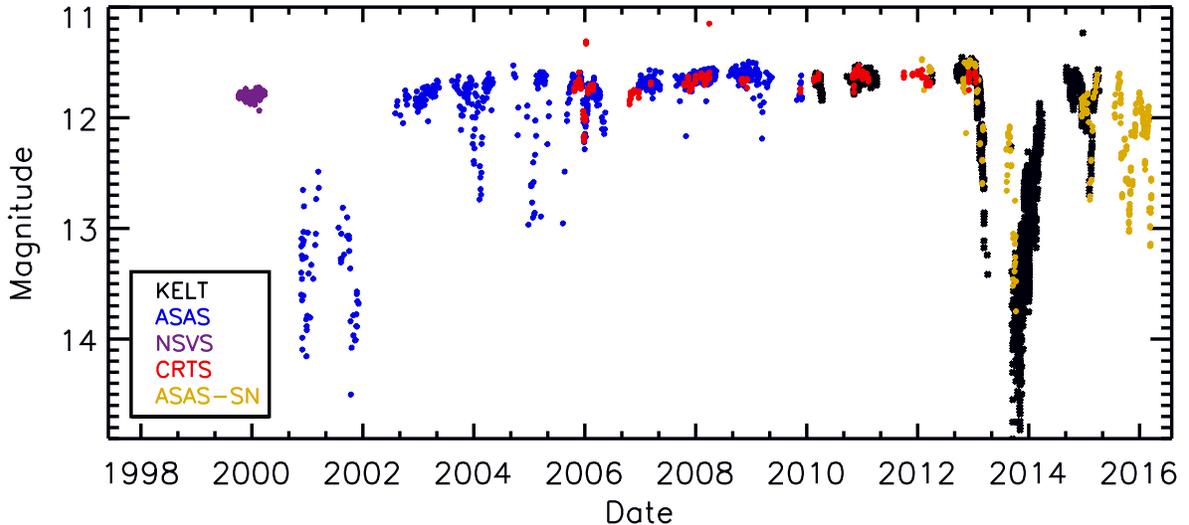,clip=,width=0.88\linewidth}
\caption{The KELT-South (Black), ASAS (Blue), NSVS (Violet), CRTS (Red), and ASAS-SN (Yellow) observations plotted from 1999 to 2016. We have applied a vertical offset to the KELT-South, ASAS, CRTS, and NSVS data to match the quiescent magnitude seen in the ASAS-SN $V$-band data. However, there has been no attempt to place all the data on the same absolute scale.}
\label{figure:FullLC}
\end{figure*}

In addition to the long duration variability (months to years) observed in these disk eclipsing systems, YSOs are also known to display shorter duration (days to weeks) photometric variability caused by processes near the host star, such as accretion, hot spots, circumstellar extinction, and inner disk instabilities \citep{Herbst:1994, Stassun:1999, Bouvier:2007}. The host star's mass, magnetic field, and rotation rate can influence the amplitude, periodicity, and duration of this variability \citep{Herbst:1999, Grankin:2007, Cauley:2012}. Also, the architecture of the circumstellar disk can contribute to the variability \citep{Flaherty:2012}. The Young Stellar Object Variability (YSOVAR) project \citep{Morales:2011, Rebull:2014} is conducting a high-cadence survey of known young clusters over a wide range of wavelengths to better understand how these specific system parameters affect the observed variability.  After observing the cluster NGC 2264, they separated their light curves into unique classes of variability based on periodicity and observational characteristics. They found that the most common variability phenomena were optical fading events, called ``dippers'', that are attributed to circumstellar obscuration. This interpretation is supported by the finding that the amplitudes of the dippers observed in the infrared by Spitzer are less than the amplitudes seen in the optical. Along with the dippers, YSOVAR identified stars that would quickly brighten and fade in $<$ 1 day, which they called ``bursts''. The burst phenomenon is believed to be caused by accretion instabilities. The analysis of NGC 2264 has suggested multiple origins to explain the range of variability observed in both the optical and infrared \citep{Cody:2014}.

In this paper, we present new results on the YSO, DM Orionis. First identified as a young stellar object candidate by \citet{Sanchez:2014}, DM Ori has the spectrum of a reddened G star and shows H$\alpha$ emission \citep{Dopita:2007}. Its properties are summarized in Table \ref{tbl:Host_Lit_Props}. The significant infrared excess indicates that DM Ori is surrounded by a very large protoplanetary disk. DM Ori underwent two large dimming events in 2000-2002 and 2013-2014, which we interpret to be caused by a structure within DM Ori's circumstellar disk. Photometric observations of DM Ori, outside the large dimming events, show shorter duration (days to weeks) dimmings consistent with the dippers found by \citet{Cody:2014} in NGC 2264. Unfortunately, we do not have color information during either dimming event to look for a color reversal, but it is possible that DM Ori is a newly identified low-mass UXor star. Whether DM Ori is a UXor star or not, in this paper we conclude that it is a newly discovered disk eclipsing system.
%It has been suggested that these dips in the optical light curve are caused by enhanced dust extinction.

The paper is structured in the following way. Our observations from KELT-South, ASAS-SN, and archival data are presented in \S2. The 2000-2002 and 2013-2014 large dimming events and the observed out-of-dimming variability are discussed in \S3. We explore several interpretations for the photometric results and the likelihood of each scenario in \S4. Our results and conclusions are summarized in \S5.

%($\alpha =$ 05$^{h}$ 43$^{m}$ 39$\fs$000 $\delta =$ -05$\degr$ 04$\arcmin$ 02$\arcsec$83 J2000, \citet{Hog:2000}), $V$ = 11.72 $\pm$ 0.12
%\section{Characteristics of the DM Ori System}
%DM Orionis was spectroscopically observed by \citet{Tisserand:2013}. Using observations from WiFeS on the 2.3 meter Australian National University telescope \citep{Dopita:2007}, they found DM Ori to have a spectrum consistent with a G-type star but one that is reddened and shows H$\alpha$ emission. DM Ori was identified as a young stellar object candidate by \citet{Sanchez:2014}. For a full list of the known parameters for DM Ori, see Table \ref{tbl:Host_Lit_Props}. 
%{\bf [Is this all that is known about DM Ori?? Is it a previously known YSO/TTS? Has anyone said anything interesting about this star in the past?]} 
\begin{figure*}[!ht]
\centering\epsfig{file=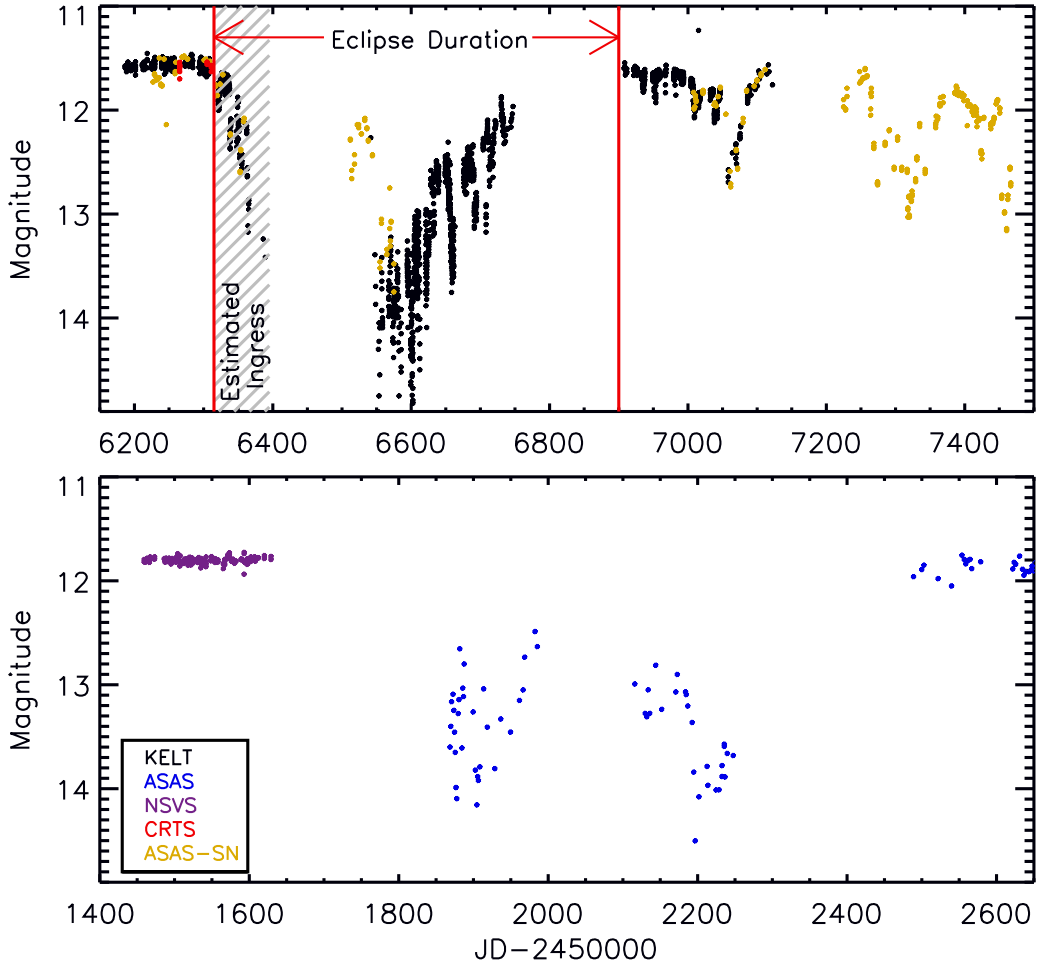,clip=,width=0.8\linewidth}
\caption{The KELT-South (Black), ASAS (Blue), NSVS (Violet), CRTS (Red), and ASAS-SN (Yellow) observations for the 2013-2014 (Top) and the 2000-2002 (Bottom) dimming events. We have applied a vertical offset to the KELT-South, ASAS, CRTS, and NSVS data to match the quiescent magnitude in each data set to the quiescent magnitude of the ASAS-SN $V$-band observations. However, there has been no attempt to place all the data on the same absolute scale.}
\label{figure:eclipse}
\end{figure*}

\section{Data}
Over the last few decades, DM Ori has been observed in a variety of photometric surveys. Below we describe the observations used in our analysis with the results shown in Figures \ref{figure:FullLC} and \ref{figure:eclipse}.

\subsection{KELT-South}
Starting in 2007, the Kilodegree Extremely Little Telescope (KELT) survey has been observing a large portion of the Southern sky with the primary objective of discovering transiting exoplanets around bright stars. Located at the South African Astronomical Observatory (SAAO) in Sutherland, South Africa, the KELT-South telescope uses a Mamiya 645-series wide-angle lens with a 42mm aperture and a 80mm focal length (f/1.9). The survey is optimized to observe stars $8 < V < 11$ with a photometric precision of $\sim$1$\%$ \citep{Pepper:2012}. Each KELT-South field is observed with a nonstandard filter equivalent to a very broad $R$-band, a $26^{\circ}$ $\times$ $26^{\circ}$ field-of-view, and a 23$\arcsec$ pixel scale. A typical KELT-South field is observed every 12-15 minutes. Located in KELT-South field 05 ($\alpha$ =  06hr 07m 48.0s, $\delta$ = $+3^{\circ}$ 00$\arcmin$ 00$\arcsec$), DM Ori was observed 2893 times from UT 2010 February 28 to UT 2015 April 09. A complete description of the data reduction process can be found in \citet{Siverd:2012} and \citet{Kuhn:2016}. The median per-point error for the entire KELT-South dataset is $\sim$0.03 mag but the source is close to the faint magnitude limit during the 2013 dimming event. 

\subsection{Northern Sky Variability Survey (NSVS)}
The Northern Sky Variability Survey (NSVS) made unfiltered observations of the Northern sky of sources with optical magnitudes from 8 to 15.5. Using four telephoto lenses as part of the Robotic Optical Transient Search Experiment (ROTSE-I), NSVS surveyed $\sim$14 million objects for one-year time spans \citep{Wozniak:2004}. The telescope is located just outside the Los Alamos National Laboratory. NSVS observed DM Ori from UT 1999 October 07 to UT 2000 March 25 obtaining 198 observations. The typical point-to-point photometric scatter is $\sim$0.02 mag. 
%\subsection{American Association of Variable Star Observers (AAVSO)}

%\subsection{Digital Access to a Sky Century at Harvard (DASCH)}

\subsection{Catalina Real-time Transient Survey (CRTS)}
The Catalina Real-time Transient Survey is designed to survey a large portion of the sky (33,000 Deg$^{2}$) to discover rare transient phenomena. The survey uses the 1.5m telescope at Mt. Lemmon, the 0.7m Catalina Sky Survey Schmidt telescope at Mt. Bigelow, and the 0.5m Schmidt survey telescope at Siding Springs. The observations are made publicly available within minutes of begin taken. For a more detailed description of the observing procedure, see \citet{Drake:2009}. DM Ori was observed 195 times from UT 2005 October 24 to UT 2013 January 18 with a typical photometric error of 0.05 mag.

\subsection{All-Sky Automated Survey (ASAS)}
The All-Sky Automated Survey (ASAS) is a project dedicated to observing all stars brighter than 14 magnitude over the entire sky. The project was designed to find and catalog variable stars. The survey has two observing locations, Las Campanas, Chile and Haleakala, Maui and observes simultaneously in the $V$ and $I$ bands. Each station has two wide-field Minolta 200/2.8 APO-G telephoto lenses with a 2K$\times$2K Apogee CCD corresponding to a 8.8$^{\circ}\times8.8^{\circ}$ field-of-view. ASAS observed DM Ori 654 times in the $V$ band from UT 2000 November 20 to UT 2009 December 01. The photometric error ranges from 0.03 to 0.1 mag. See \citet{Pojamanski:1997} for a detailed description of the survey and data reduction process.

\subsection{All-Sky Automated Survey for SuperNovae (ASAS-SN)}
With the objective of monitoring the entire sky down to $V \sim 17$ mag, the All-Sky Automated Survey for SuperNovae (ASAS-SN, \citet{Shappee:2014}) project has been focused on discovering nearby supernovae. The survey is comprised of two units containing four 14-cm aperture Nikon telephoto lenses each, with one located on Mount Haleakala in Hawaii and the other at the Cerro Tololo InterAmerican Observatory (CTIO) in Chile. The telescopes are hosted by the Las Cumbres Observatory Global Telescope Network \citep{Brown:2013}. Each lens has a 2k $\times$ 2k thinned CCD, corresponding to a $4.5\times4.5$ degree field-of-view and a 7$\farcs$8 pixel scale. Together, the two units can cover the entire visible sky every two days. ASAS-SN observed the field containing DM Ori 332 times from UT 2012 January 23 until UT 2016 March 19. The median photometric error for DM Ori is $\sim$0.1 mag.

%\subsection{Broadband Photometry from the Literature for Spectral Energy Distribution Modeling}
%To better understand the circumstellar environment of DM Ori and help constrain the occulting body's physical parameters, we assembled all available photometric data in the literature (summerized in Table \ref{tbl:Host_Lit_Props}). We then modeled the SED of the system. A description of our model and analysis is presented in \S\ref{sec:SED}.

\section{Results}
In this section we discuss the photometric variability observed for DM Ori using the KELT-South, NSVS, CRTS, ASAS, and ASAS-SN photometric observations. Specifically, we focus our discussion on the 2000-2002 and 2013-2014 large dimming events. In section 5 we discuss possible explanations for the large dimmings. 

\subsection{Out of Dimming Variability}
Young T Tauri stars can show a wide range of photometric variability due to accretion and/or circumstellar extinction  \citep{Herbst:1994}. In some cases, the accretion-driven variability can be periodic in nature, since the accretion will create ``hot spots'' on the stellar surface that periodically cross our line of sight due to stellar rotation. Using the Lomb-Scargle periodicity search algorithm \citep{Lomb:1976, Scargle:1982, Press:1989} in the VARTOOLS package \citep{Hartman:2012}, we searched for periodic signals in the photometric data. Unfortunately, there are no publicly available photometric observations of DM Ori prior to 1999. Our observations of DM Ori, outside the 2000-2002 and 2013-2014 large dimming events, display shallow (0.1 -- 0.5 mag) and short duration ($<$ 1 month) dimmings that are similar to the optical dippers found by \citet{Cody:2014} in NGC 2264. Prior to the 2000-2002 eclipse, we have $\sim$6 months of observations from NSVS. We find no significant periodic variability in the NSVS observations but we do see non-periodic photometric variability with a peak-to-peak amplitude of 0.1 mag. Similarly, we find no significant periodic variability in the full KELT-South, ASAS-SN, CRTS, or ASAS data sets. Even when we remove the large dimming events and only look at individual observing seasons during periods of relative quiescence, we find no significant periodic variability. 

Interestingly, after both the 2000-2002 and 2013-2014 large dimming events, we see large amplitude variability. This variability was observed by KELT-South and ASAS-SN after the 2013-2014 dimming and by ASAS after the 2000-2002 dimming. These smaller dimming features last for $\sim$60-80 days and have a peak-to-peak amplitude of $\sim$1 mag. This out of dimming variability occurs immediately after the 2013-2014 dimming while there is a $\sim$1 year duration quiescent period after the 2000-2002 event before the variability begins. In \S\ref{sec:interp} we discuss the possibility that these are trailing features associated with the occulting mechanism. 

\subsection{2000-2002 Dimming}
In late 2000 through early 2002, DM Ori was $\sim$1.5 mag fainter than its quiescent magnitude of $V\sim$11.7 (See Table \ref{tbl:Host_Lit_Props} and Figure \ref{figure:eclipse}(a)). Unfortunately, our photometric coverage of the 2000-2002 dimming is quite poor since the dimming began after the last NSVS observations, on UT 2000 March 25. The first ASAS observation on UT 2000 November 20 occurred during the dimming event. In addition, the 2002 seasonal gap occurred while DM Ori was returning to its normal brightness. Therefore, we have no coverage of the ingress or egress for the 2000-2002 dimming. We can place an upper limit on the dimming duration of $\sim$860 days, the time from the last NSVS observation on UT 2000 March 25 to the first observation by ASAS after the 2002 seasonal observing gap, UT 2002 August 02. Also, we can place a lower limit on the duration to be $>$396 days, the total duration of the ASAS observations during the dimming. We observed $\sim$1 mag peak-to-peak variability on a timescale of 60-80 days after the end of the primary event. If the dimming is caused by an orbiting body or cloud of material, the similarities of the photometric variability during and after the dimming would suggest they are associated with each other. We explore this possibility more in \S\ref{sec:interp}.

\subsection{2013-2014 Dimming}
Near the end of UT 2013 January, the DM Ori system began to dim by $\sim$1.5 mag over an $\sim$80 day period (ingress timescale, See Figure \ref{figure:eclipse}(b)). Interestingly, this timescale is consistent with the post-dimming variability observed after both the 2000-2002 and 2013-2014 events. We estimate the duration of this eclipse to be $<$585 days, but this is likely an overestimate since the egress of the event occurred during the 2014 seasonal observing gap. The lower limit for the eclipse duration is $>$420 days. Since the 2000-2002 dimming is $<$860 days and not well constrained, it is possible that both the 2000-2002 and 2013-2014 dimming events are similar in duration. The estimated centers of each event are separated by $\sim$12.5 years. Therefore, it is plausible that we are seeing two instances of a $\sim$1.5 mag, $<$585 day long dimming event, with a period of $\sim$12.5 years.

During the second eclipse, we again observed a large amount of structure. Unfortunately, DM Ori at minimum is near the magnitude limit of KELT-South, limiting our analysis. When combining the KELT-South and ASAS-SN observations, we can see that DM Ori was V$\sim$14 at its dimmest point. During the eclipse, DM Ori returned to V$\sim$12 for about $\sim$30 days starting at JD = 2456511 (TT), before fading back to V$\sim$14 on a similar timescale to the ingress. Unlike the initial ingress, which lasted $\sim$80 days, the brightening beginning at JD$\sim$2456590 (TT) is gradual. Without coverage of the end of the 2013-2014 dimming event, we place an upper limit on the brightening duration of $\sim$310 days. During this long gradual brightening, DM Ori's brightness fluctuates with a peak-to-peak amplitude of $<$1 mag. Specifically, from JD$\sim$2456650 (TT) to the last KELT-South observation prior to the 2014 season observing gap (JD = 2456747.2058 (TT)), there is large amount of variability on a timescale of $\sim$20 days with decreasing amplitude. 

\subsection{SED Analysis and Implications}
\label{sec:SED}

\begin{figure}[!ht]
%\centering\epsfig{file=dmori_sed_fin.pdf,angle=0,clip,trim=100 50 70 70,width=\linewidth}
\centering\epsfig{file=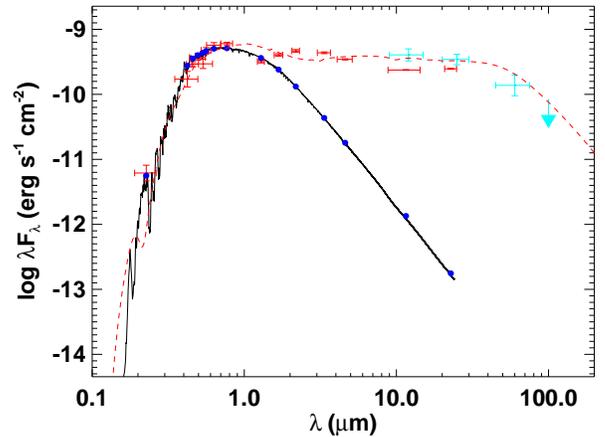,angle=-90,clip,trim=80 50 70 70, width=\linewidth}
\caption{The Spectral Energy Distribution of DM Ori. The red points are the GALEX to WISE photometry, and the cyan points are the archival IRAS photometry and upper limits showing the large IR excess. The solid curve and blue points are the best fitting NextGen stellar atmosphere model for the photometry from 0.2--1.2 $\mu$m. The red dashed curve is an example of a disk model that explains the IR emission.}
\label{SED_Fit}
\end{figure}

We assembled photometry for DM Ori from GALEX, Tycho, APASS, 2MASS, and WISE, providing a well-sampled SED over the wavelength range 0.2--22 $\mu$m. We also used archival fluxes from IRAS at 12--100 $\mu$m (see Table \ref{tbl:Host_Lit_Props}). Even without a fit to the observed SED, a very large infrared excess is obvious (Fig.~\ref{SED_Fit}). 

To estimate the underlying photospheric properties, we fit NextGen stellar atmospheres to only the UV and optical measurements. 
The resulting best-fit stellar atmosphere model has the following properties: $T_{\rm eff} = 6740 \pm 250$ K, [Fe/H] = $0.0\pm0.2$, and $A_V = 1.1\pm0.3$. 

\begin{figure}[!ht]
\centering\epsfig{file=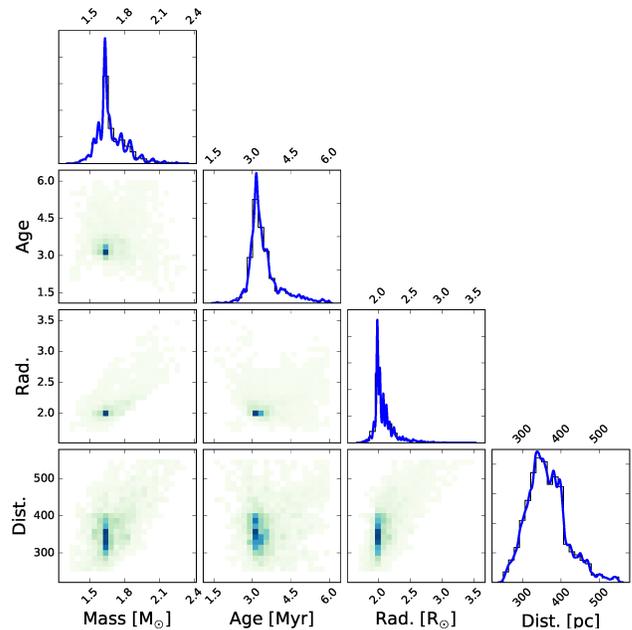,trim={0 1.5in 0 1.2in},clip=,angle=-0,width=1.0\linewidth}
\caption{Posterior distributions for mass, age, radius, and distance of DM Ori inferred from the Dartmouth Stellar Evolution Models. The color-scale in each panel indicates areas of higher posterior probability. Along the diagonal are plotted the marginalized distributions for each parameter and their corresponding kernel density estimation (blue line).}
\label{Radius_MCMC}
\end{figure}

In order to determine the fundamental parameters of DM Ori, we take a similar approach to our V409 Tau study \citep{Rodriguez:2015}, namely, to infer stellar parameters using stellar evolution isochrone models. However, unlike the V409 Tau analysis, we are very limited in the available data for DM Ori. We are only able to use the temperature of \teff$=$6740$\pm$250 K determined from the fits to the SED, as well as the star's B- and V-band magnitudes (See Table \ref{tbl:Host_Lit_Props}). The longer wavelength photometric data for DM Ori are not useful due to contamination at these wavelengths from the disk. Assuming DM Ori is associated with the OB1c or the Orion Nebula Cluster portion of the Orion association \citep{Bally:2008} allows us to place priors on some of the parameters. We assume uniform priors for distance ($250-550$ pc), metallicity ($[Fe/H] = -0.2-0.2$ dex), and age ($0-6$ Myr). We also assume an extinction of $A_{V} = 1.076$ based on our best-fit SED. We use both the Dartmouth Stellar Evolution Models \citep{Dotter:2008} and the MIST stellar evolution models \citep{Choi:2016} combined with a Markov-Chain Monte Carlo (MCMC) to derive posterior distributions for the mass, age, radius, and distance of DM Ori (see Figure \ref{Radius_MCMC}). From this analysis, we determine the most probable parameter values based on the posterior medians with errors from their 68\% confidence interval uncertainties. For the Dartmouth models we find a mass of 1.64$^{+0.16}_{-0.07}$ $M_{\odot}$, an age of 3.29$^{+0.64}_{-0.28}$ Myr, a radius of 2.03$^{+0.17}_{-0.05}$ $R_{\odot}$, and a distance of 357$^{+49}_{-44}$ pc. Similarly, for the MIST models, we find a mass of 1.49$^{+0.30}_{-0.09}$ $M_{\odot}$, an age of 3.07$^{+1.34}_{-1.21}$ Myr, a radius of 2.05$^{+0.15}_{-0.13}$ $R_{\odot}$, and a distance of 385$^{+59}_{-41}$ pc. Both sets of stellar evolution models produce consistent parameters for DM Ori, and the inferred age and distance are consistent with it being associated with the Orion star-forming region. To rule out the possibility that DM Ori is actually a foreground star in our line-of-sight to the Orion association, we reran our analysis with distance prior of 0-550pc. Our results are very similar to the distribution shown in Figure \ref{Radius_MCMC}, providing strong evidence that DM Ori is not a foreground star with a distance $<$ 250 pc.

Our interpretation in the next section of the large dimming events observed in the DM Ori light curve involves an occultation of the star by a disk that is optically thick well into the far-infrared. The prominent infrared excess in the SED strongly suggests the presence of such a disk. We generated a model SED of a star plus disk following the procedure of \citet{Jensen:1997}. This is a simple analytical model of a geometrically thin disk with a power-law surface density and power-law temperature profile. To obtain a representative fit, we varied only the disk temperature profile exponent, the total disk mass, the disk luminosity, and the size of the inner disk hole. We adopted the stellar $T_{\rm eff}$ and $A_V$ from above for the blue side of the SED. The resulting model is shown in Fig.~\ref{SED_Fit}, with a disk mass of 0.02 M$_\odot$, a disk luminosity of 2 L$_\odot$, an inner disk hole radius of 0.055 AU, and a disk temperature power-law exponent of 0.5. We emphasize that this not intended to be a proper fit of the data, but simply to illustrate the type of massive, optically thick disk that is required to reasonably reproduce the observed SED. 

We attempted to constrain the inclination of DM Ori`s circumstellar disk with SED modeling using the self-consistent irradiated, accretion disk models of \citet{D'Alessio:2006}.  We refer the reader to \citet{Espaillat:2015} and references within for more details on the model and modeling procedure. We created a model grid of 1,080 disks using stellar parameters of M$_{*}$=1.5{\msun}, T$_{*}$=6740K, R$_{*}$=2.1{\rsun}, and a distance of 385 pc.  To the best of our knowledge, there is no measurement of the accretion rate in DM Ori and so we adopt a mass accretion rate of 1 $\times$ 10$^{-8}$ \msun yr$^{-1}$, which is the average accretion rate of T Tauri stars (Hartmann et al. 1998).  We also adopt a dust sublimation temperature of 1400 K to set the inner radius of the disk. We varied the disk outer radius (50 AU, 150 AU, and 250 AU), viscosity ($\alpha$ = 0.1, 0.01, and 0.001), and dust settling parameters ($\epsilon$ = 1.0, 0.1, 0.01, 0.001). The minimum grain size in the disk atmosphere is held fixed at 0.005 $\micron$ while we varied the maximum grain size (0.05, 0.25, 1.0, 5.0, and 100.0 $\micron$. We then tried disk inclinations of 10, 20, 40, 50, 60, and 80 degrees. Unfortunately, we were not able to reproduce the SED or constrain the diskÕs inclination.  This may be the result of the intrinsic photometric variability of DM Ori or some physical structure in the disk that the models do not account for.

%\subsection{Determining Stellar Parameters}

\section{Interpretation and Discussion}
\label{sec:interp}
Given observations showing only two extended dimming events, we cannot definitively establish that this is a strictly periodic phenomenon. However, this interpretation is consistent with the behavior of V409 Tau and AA Tau, which have periodically recurring occultations of the host star by a feature in their protoplanetary disk. For DM Ori, the observations suggest a period of $\sim$12.5 years. This scenario is supported by the similarities between the observed 2000-2002 and 2013-2014 dimming events, as well as the post-dimming variability. This explanation is dependent on the inclination of the DM Ori disk, a presently unknown parameter. 
%In this section we explore the most plausible interpretation for the two observed dimming events of DM Ori, specifically that DM Ori is being periodically occulted by a portion of the very large disk. This scenario is supported by the similarities between the observed 2000-2002 and 2013-2014 dimming events, as well as the post-dimming variability. This explanation is dependent on the inclination of the DM Ori disk, a presently unknown parameter. 

We model the dimmings as occultations of the host star by a large body in Keplerian motion. In this model, the occulting body can have a leading edge perpendicular to its direction of motion (rectangular shape) or a leading edge inclined to its direction of motion (``wedge-shaped,'' see Figure 9 from \citet{Rodriguez:2015}). Using the estimated ingress timescale of $\sim$80 days, we calculate the transverse velocity of the occulter to be V$_T \sim 2R_{\star}/(T_{ingress}\sin\theta)$. Without an estimate of the DM Ori disk radius, we constrain the occulting body to be located within 215 AU, the radius of the disk around AA Tau \citep{Kitamura:2002}. Setting $\theta$ = 0 (leading edge perpendicular to the direction of motion), we get a transverse velocity of 412.6 \ms.  Using this velocity, and the total duration of the 2013-2014 event of 585 days, this would imply that the occulting body is $\sim$0.14 AU wide. 

Assuming Keplerian motion, we can determine the semimajor axis of the orbit as

\begin{equation}
a = \frac{M_\star GT_{\rm ingress}^2}{4R_\star^2}.
\end{equation}

\noindent Adopting the estimated transverse velocity and ingress time scale would require the occulting body to be located far outside the extent of DM Ori's disk (a$\sim$7700 AU), making this scenario implausible. To constrain the occulting body to be at $\sim$215 AU (outer edge of the disk) with an 80 day ingress, requires a wedge angle of $<$9.6$^{\circ}$, almost parallel to the direction of motion. If the occulter is located at the edge of the disk, it would have a transverse velocity of $\sim$2.5 \kms\ and be $\sim$0.84 AU wide. 

Since both dimming events are similar in depth and duration, it is possible that one feature caused both the 2000-2002 and 2013-2014 dimming events. This scenario would imply that the occulting body is orbiting DM Ori with a period of $\sim$12.5 years (the time between the two events). A period of $\sim$12.5 years implies a semi-major axis for the occulting feature of $\sim$6.1 AU. For the same wedge-shaped model, this would require the leading edge of the occulting body to be at an angle of $\sim$1.6$^{\circ}$. Using this semi-major axis and assuming Keplerian motion, the occulting body would be moving $\sim$14.6 \kms\ and be $\sim$4.9 AU in width to cause the dimming event observed in 2013-2014. 

If the dimmings are caused by a dust cloud or a warp in the disk, the post-dimming photometric variability observed after both events may be caused by trailing circumstellar material in a similar orbit to the main occulting body. Interestingly, the estimated semi-major axis and dimming duration ($<$585 days) would be consistent with the heating and cooling timescales for the inner disk (where a$<$10 AU and duration $<$10 years) found by \citet{Nelson:2000}. However, the required scale height to cross our line-of-sight is directly related to the unknown disk inclination. It is likely that the in-eclipse structure during the both large dimming events is caused by gaps or opacity changes in the occulting body. 

The interpretation of these dimming events as being caused by a feature (possibly a warp or dust clump) in the disk is dependent on knowing the host star's mass and the inclination of the disk. For a disk orientation close to edge-on, the interpretation that the dimming mechanism is located in the disk is very plausible. As first proposed by \citet{Kenyon:1987}, the scale height of protoplanetary disks can increase with radius, and this has been observed for disks in the Orion association \citep{Smith:2005}. Therefore, a nearly edge-on disk ($>$70$^{\circ}$), would only require a small perturbation to cross our line of sight. For systems with a non-edge-on disk architecture, the dimming mechanism becomes more complex. One scenario is that a screen of hot dust can be created in the inner disk and be pushed across our line-of-sight by high disk winds \citep{Petrov:2015}. A second scenario is that the disk grains are very fine, remaining mixed with the gas and dispersed throughout the system (and not constrained to the disk plane). These grains can then easily clump up (out of the disk plane) and cause large photometric dimming events for systems with inclinations between 45$^{\circ}$ and 68$^{\circ}$ \citep{Natta:2000}. 

%\subsection{Favored Interpretation: }
%\subsection{Alternate Explanations}

\section{Summary and Conclusions}
From studying disk-eclipsing systems, where a young star is occulted by a feature in its circumstellar environment, we can learn about the diversity of environments surrounding young stars and how each scenario can influence planet formation within the system. These systems could provide insight into the formation and evolution of a protoplanetary disk, explaining the large range of exoplanetary architectures. 

Observations of DM Ori by the KELT-South, ASAS, NSVS,  CRTS, and ASAS-SN surveys show two separate $<$ 585 day long, $\sim$1.5 mag depth dimming events in 2000-2002 and 2013-2014. We estimate the ingress timescale to be $\sim$80 days. To determine key characteristics of the occulting body, we apply a simple model in which the dimming is caused by an occultation of the host star by a large opaque body that has a leading edge inclined to its direction of motion. Combining our estimated depth, duration, and ingress timescale, we estimate that the occulting body would be moving at $\sim$14.6 \kms, be $\sim$4.9 AU in width, and would be $>$6.1 AU from the star, assuming Keplerian motion. Our SED analysis suggests the presence of a very large disk but we do not have a constraint on the disk inclination. Unfortunately, the inclination of the disk is also crucial to understanding the specific dimming mechanism. If the disk is close to edge-on ($>$70$^{\circ}$), then these dimming events may be caused by a warp, perturbation, or dust clump orbiting within the disk plane. If the disk is inclined to our line-of-sight ($<$70$^{\circ}$), then it is possible that the dimming events were caused by heating and cooling of the inner disk causing the disk grains to fluctuate in their spatial distribution \citep{Nelson:2000}. It is also possible that disk grains are being pushed across the stellar photosphere by high disk winds or by grains still mixed in with the gas \citep{Natta:2000, Petrov:2015}.  

The discovery of the two large dimming events of DM Ori are a strong motivation for continued photometric monitoring, especially during the next predicted dimming in July of 2026. Specifically, future observations should look for the color reversal, a characteristic of a late-type UXor star. Millimeter mapping of the DM Ori system should be able to detect the disk and measure its inclination. These observations would provide additional evidence for the specific mechanism causing the 2000-2002 and 2013-2014 large dimming events. Future surveys like the Large Synoptic Survey Telescope (LSST) should increase the number of disk eclipsing systems known by at least an order of magnitude, providing a large sample with which to study these rare systems and their potential value for understanding planetary formation. 

%DM Ori appears to be a rare but potentially important example of the signatures of planet formation in a young star's protoplanetary disk, consistent with some theories of planet formation. 

\acknowledgments
Early work on KELT-North was supported by NASA Grant NNG04GO70G. J.A.P. and K.G.S. acknowledge support from the Vanderbilt Office of the Provost through the Vanderbilt Initiative in Data-intensive Astrophysics. This work has made use of NASA's Astrophysics Data System and the SIMBAD database operated at CDS, Strasbourg, France.

Work by B.S.G. was partially supported by NSF CAREER Grant AST-1056524. Work by K.G.S. was supported by NSF PAARE grant AST-1358862. B.J.S. is supported by NASA through Hubble Fellowship grant HF-51348.001 awarded by the Space Telescope Science Institute, which is operated by the Association of Universities for Research in Astronomy, Inc., for NASA, under contract NAS 5-26555.  CSK and KZS are supported by NSF grants AST-1515876 and AST-1515927. TW-SH is supported by the DOE Computational Science Graduate Fellowship, grant number DE-FG02-97ER25308. P.C. is supported by the NASA grant NNX13AI46G. 

The CSS survey is funded by the National Aeronautics and Space Administration under Grant No. NNG05GF22G issued through the Science Mission Directorate Near-Earth Objects Observations Program. The CRTS survey is supported by the U.S.~National Science Foundation under grants AST-0909182 and AST-1313422.

Development of ASAS-SN has been supported by NSF grant AST-0908816 and CCAPP at the Ohio State University.  ASAS-SN is supported by NSF grant AST-1515927, the Center for Cosmology and AstroParticle Physics (CCAPP) at OSU, the Mt. Cuba Astronomical Foundation, George Skestos, and the Robert Martin Ayers Sciences Fund.

\bibliographystyle{apj}

\bibliography{DM_Ori}

\end{document}